\documentclass[aoas,MSNbibl,nameyear,dvips]{arximspdf}
\usepackage{graphicx}
\doi{10.1214/13-AOAS645} 
\volume{7}
\issue{3}
\pubyear{2013}
\firstpage{1311}
\lastpage{1333}

\makeatletter

\newcommand{\implies}{\Longrightarrow}

\def\R{{{\mathbb{R}}}}

\makeatother

\begin{document}
\begin{frontmatter}

\title{Learning a nonlinear dynamical system model of~gene regulation:
A perturbed steady-state approach}
\runtitle{Learning gene network from steady-state}

\begin{aug}
\author[A]{\fnms{Arwen} \snm{Meister}\ead[label=e1]{arwenb@stanford.edu}},
\author[A]{\fnms{Ye Henry} \snm{Li}\thanksref{t1}\ead[label=e2]{ywli@stanford.edu}},
\author[A]{\fnms{Bokyung} \snm{Choi}\ead[label=e3]{bkchoi@stanford.edu}}
\and\\
\author[A]{\fnms{Wing Hung} \snm{Wong}\corref{}\thanksref{t2}\ead[label=e4]{whwong@stanford.edu}}
\runauthor{Meister, Li, Choi and Wong}
\affiliation{Stanford University}
\address[A]{Department of Statistics\\
Stanford University\\
Sequoia Hall\\
390 Serra Mall\\
USA\\
\printead{e1}\\
\hphantom{E-mail: }\printead*{e2}\\
\hphantom{E-mail: }\printead*{e3}\\
\hphantom{E-mail: }\printead*{e4}} 
\end{aug}

\thankstext{t1}{Supported in part by NIH Grant GM007276.}

\thankstext{t2}{Supported in part by NIH Grant R01-HG006018 and NSF Grant DMS-09-06044.}

\received{\smonth{7} \syear{2012}}
\revised{\smonth{3} \syear{2013}}

%
\begin{abstract}
Biological structure and function depend on complex regulatory
interactions between many genes. A wealth of gene expression data is
available from high-throughput genome-wide measurement technologies,
but effective gene regulatory network inference methods are still
needed. Model-based methods founded on quantitative descriptions of
gene regulation are among the most promising, but many such methods
rely on simple, local models or on ad hoc inference approaches lacking
experimental interpretability. We propose an experimental design and
develop an associated statistical method for inferring a gene network
by learning a standard quantitative, interpretable, predictive,
biophysics-based ordinary differential equation model of gene
regulation. We fit the model parameters using gene expression
measurements from perturbed steady-states of the system, like those
following overexpression or knockdown experiments. Although the
original model is nonlinear, our design allows us to transform it into
a convex optimization problem by restricting attention to steady-states
and using the lasso for parameter selection. Here, we describe the
model and inference algorithm and apply them to a synthetic six-gene
system, demonstrating that the model is detailed and flexible enough to
account for activation and repression as well as synergistic and
self-regulation, and the algorithm can efficiently and accurately
recover the parameters used to generate the data.\looseness=-1
\end{abstract}

%

\end{frontmatter}

\section*{Introduction}
Complex interactions between many genes give rise to the biological
structure and function that sustain life. The Central Dogma [\citet
{jacob61,crick70}]\vadjust{\goodbreak} provides a qualitative description of how these
processes occur, but precise quantitative modeling is still needed
[\citet{tyson03,rosenfeld11}]. Research into the detailed mechanisms of
gene expression over the past few decades has shown that expression is
regulated by a complex system of gene interactions. Recently,
microarray and sequencing technologies [\citet{derisi97,ren00,robertson07,mortazavi08}] have enabled high-throughput genome-wide
expression level measurements. This data enables detailed study of gene
networks [\citet{holstege98,lee02,tegner03,segal03,bar-joseph03,hu07,zhou07}]. The goal is to understand how genes interact to give
rise to the biochemical complexity that allows organisms to live, grow
and reproduce.

Gene expression measurements contain information useful for
reconstructing the underlying interaction structure [\citet{derisi97,holstege98,hughes00}] because gene regulatory systems have a defined
ordering [\citet{avery92}], forming pathways that connect to form
networks [\citet{alon07,desmet10}]. Many gene regulation pathways have
been discovered over the past few decades [\citet{hartwell10,alberts07}]. At the turn of the century, researchers began applying
statistical tools to genome-wide expression data to understand complex
gene interactions. Eisen et al. showed that genes from the same
pathways and with similar functions cluster together by expression
pattern [\citet{eisen98}]. Soon afterward, module-based network inference methods
appeared, which group co-expressed genes into cellular function modules
[\citet{segal03,bar-joseph03}]. Recently, methods based on descriptive
but nonmechanistic mathematical models [\citet{gardner03,tegner03,bansal07,faith07,friedman04}] have gained prominence. These models
describe gene regulation quantitatively and can be used to simulate and
predict systems behaviors [\citet{palsson11,dehmer11}]. However, more
work is needed to develop effective model-based methods for inferring
gene network structure from experimental data.

Existing inference methods typically rely either on heuristic
approaches or on very simple, local models, like linear differential
equation models in a neighborhood of a particular steady-state.
Statistical corrrelation is a common method of establishing network
connections [\citet{dehmer11}] and can be very useful when hundreds or
thousands of genes are monitored under specific, local cellular
conditions (e.g., for grouping genes with similar functions).
However, this approach works poorly when perturbations drive the
network far from the original steady-state. Global nonlinear models are
essential to account for complex global system behaviors, like the
transformation of a normal cell into a cancerous cell due to the
amplification of a particular gene.

As a basis for our inference approach, we chose a standard global
nonlinear model: the quantitative, experimentally interpretable
biophysics-based ordinary differential equation (ODE) gene regulation
model of Bintu et al. (\citeyear{bintu05models,bintu05apps}). Many
models of this type have been proposed, and the idea traces back to the
beginning of systems biology in the biophysics field
[\citet{ackers82,shea85,vonhippel74}], but the Bintu model is
widely accepted within the biophysics community
[\citet{bintu05models}]. The Bintu model is based on the
thermodynamics of RNA transcription, the process at the core of gene
expression regulation [\citet{holstege98,hu07}]. Transcription
occurs when RNA polymerase (RNAP) binds the gene promoter;
transcription factors (TFs) can modulate the RNAP binding energy to
activate or repress transcription. RNA transcripts are then translated
into protein. Bintu models the mechanism of transcription in detail,
using physically interpretable parameters. The form of the equations is
rich and flexible enough to include the full range of gene regulatory
behavior. Another notable biophysics-based model is that of the annual
DREAM competition, but it has many biochemical assumptions and model
parameters, like the Hill coefficient of transcription factor binding
events, that cannot be estimated using gene expression measurements, so
the network reconstruction requires ad hoc inference methods to learn
the underlying gene interactions
[\citet{yip10,pinna10,marbach10,schaffter11}]. Compared to the
DREAM model, the Bintu model has the advantages of simplicity and
interpretability, and better lends itself to principled inference.

In this paper, we propose an experimental design and associated
statistical method for inferring an unknown gene network by fitting the
ODE-based Bintu gene regulation model. The required data is gene
expression measurements at a set of perturbed steady-states induced by
gene knockdown and overexpression [\citet{Huang05}]. We show how to
design a sequence of experiments to collect the data and how to use it
to fit the parameters of the Bintu model, leading to a set of ODEs that
quantitatively characterize the regulatory network. Although the
original fitting problem is nonlinear, we can transform it into a
convex optimization problem by restricting our attention to
steady-states. We use the lasso [\citet{tibs96}] for parameter
selection. As a proof of principle, we test the method on a simulated
embryonic stem cell (ESC) transcription network [\citet{chickarmane08}]
given by a system of ODEs based on the Bintu model. Here, we
demonstrate that the inference algorithm is computationally efficient,
accounts for synergistic regulation and self-regulation, and correctly
recovers the parameters used to generate the data. Furthermore, the
method requires only a set of steady-state gene expression
measurements. Experimental researchers in the biological sciences can
use this method to infer gene networks in a much more principled,
detailed manner than earlier approaches allowed.

\section*{Dynamical systems model}
We model gene expression regulation as a dynamical system. Let $x \in\R
^n$ represent RNA concentrations and $y \in\R^n$ represent protein
concentrations corresponding to a set of $n$ genes. We assume that the
production rate of the RNA transcript $x_i$ of gene $i$ is proportional
to the probability $f(y)$ that RNA polymerase (RNAP) is bound to the
promoter. That is, we assume that RNA transcription occurs at a rate
$\tau_i$ whenever RNAP is bound to the promoter. We model the
probability that RNAP is bound to the promoter as a nonlinear function
$f$ of $y$, since RNAP binding is regulated by a set of TFs. Further,
we assume that the production rate of protein product $y_i$ of gene $i$
is proportional to the concentration of the RNA transcript $x_i$, and
that both the RNA transcript and protein products of gene $i$ degrade
at fixed rates ($\lambda_i^{\mathrm{RNA}}$, $\lambda_i^{\mathrm{Protein}}$),
%
\begin{eqnarray}\label{eqexpression}
\frac{dx_i}{dt} &=& \tau_i f_i(y) -
\lambda_i^{\mathrm{RNA}} x_i,
\nonumber\\[-8pt]\\[-8pt]
\frac{dy_i}{dt} &=& r_ix_i - \lambda_i^{\mathrm{Protein}}
y_i.\nonumber
\end{eqnarray}
Based on the thermodynamics of RNAP and TF binding, one can deduce the
following form for $f_i$ [Bintu et~al. (\citeyear{bintu05models,bintu05apps})]:
%
\begin{equation} \label{eqf}
f_i(y) = \frac{b_{i0} + \sum_{j=1}^m b_{ij} \Pi_{k \in S_{ij}} y_k}{1
+ \sum_{j=1}^m c_{ij} \Pi_{k \in S_{ij}} y_k},
\end{equation}
where $S_{ij}$ lists the gene products that interact to form a
regulatory complex, and $b_{ij}, c_{ij}$ are nonnegative coefficients
that must satisfy $c_{ij} \ge b_{ij} \ge0$. (We assume that the
concentration of each complex is proportional to the product of the
concentrations of the constituent proteins, and absorb the
proportionality constant into corresponding coefficients $b_{ij},
c_{ij}$.) The coefficients $b_{ij}$ and $c_{ij}$ depend on the binding
energies of regulator complexes to the promoter. $b_{i0}$ and $c_{i0}$
correspond to the case when the promoter is not bound by any regulator
($\Pi_{k \in S_{i0}} y_k = 1$), and the coefficients are normalized so
that $c_{i0} = 1$. Details and a derivation are given in the \hyperref[app]{Appendix}.

The form of $f_i$ allows us to model the full spectrum of regulatory
behavior in quantitative detail. Terms that appear in the denominator
only are repressors, and the degree of repression depends on the
magnitude of the coefficient, while terms that appear in the numerator
and denominator may act as either activators or repressors depending on
the relative magnitudes of the coefficients and the current gene
expression levels. Terms may represent either single genes or gene
complexes. The model can even be extended to account for environmental
factors that affect gene regulation, though we will not discuss it
further here.

\begin{figure}

\includegraphics{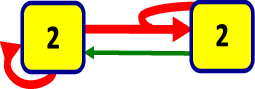}

\caption{Simple two-gene network example described by equation
(\protect\ref{eqmodelexample}) (with parameters $b_{11} = c_{11} =
0.1$ for activators; $c_{12} = 10$ for repressors; and $b_{10} = 0.01$
for constants in the numerator). Gene 1 is activated by the protein
product of gene 2 and repressed by its own product (an example of
self-regulation). Gene 2 is repressed by a complex formed by the
product of gene 1 and its own product (synergistic self-regulation). In
the diagram, the edge colors indicate activation (green) or repression
(red) and the edge weights indicate coefficient sizes, illustrated
above with typical sizes.} \label{figmodelexample}
\end{figure}

As an example, consider the simple two-gene network shown in
Figure~\ref{figmodelexample}. Suppose that genes 1 and 2 have RNA
concentrations $x_1, x_2$, and protein concentrations $y_1, y_2$,
respectively, and that gene 1 is activated by protein 2 and repressed
by its own product (protein 1), while gene 2 is repressed by a complex
formed by proteins 1 and 2. The situation corresponds to the following
equations:
%
\begin{eqnarray} \label{eqmodelexample}
\frac{dx_1}{dt} &=& \tau_1 \frac{b_{10} + b_{11} y_2 }{1 + c_{11}
y_2 + c_{12} y_1} -
\lambda_1^{\mathrm{RNA}} x_1,\qquad \frac{dy_1}{dt} =
r_1x_1 - \lambda_1^{\mathrm{Protein}}
y_1,
\nonumber\\[-8pt]\\[-8pt]
\frac{dx_2}{dt} &=& \tau_2 \frac{b_{20} }{1 + c_{21} y_1 y_2} - \lambda
_2^{\mathrm{RNA}} x_2,\qquad \frac{dy_2}{dt} =
r_2x_2 - \lambda_2^{\mathrm{Protein}}
y_2.\nonumber
\end{eqnarray}
In the notation above, we have $S_{11} = \{ 2 \}, S_{12} = \{ 1 \},
S_{21} = \{ 1, 2 \}$. The parameters $b_{10}, b_{11}, c_{11},\ldots$
determine the magnitude of the repression or activation. As this
example shows, the model is flexible enough to capture a wide range of
effects, including self-regulation (i.e., regulation of a gene by
its own protein product, most commonly as repression) and synergistic
regulation by protein complexes (two or more proteins bound together to
form a regulatory unit), in quantitative detail. Furthermore, the model
is predictive: if we know or can infer the coefficients in the model,
we can predict the future behavior of the system starting from any
initial condition.

\section*{Inference problem}

The model given by equations (\ref{eqexpression}) and (\ref{eqf}) fully
describes the evolution of RNA and protein levels and provides a
comprehensive, quantitative model of gene regulation, provided we know
the parameters. Unfortunately, $b_{ij},c_{ij}$ are extremely difficult
to measure, as they depend on binding energies of RNAP and TFs to the
gene promoter. The sheer number of measurements required to
characterize all possible TFs (both individual proteins and complexes)
also makes this approach infeasible. Therefore, our goal is to use a
systems level approach to fit the model using RNA expression data.
Specifically, we will assume that $\tau_i, \lambda_i^{\mathrm{RNA}}, \lambda
_i^{\mathrm{Protein}}$ are known or can be measured (if these quantities are not
available, we can simply absorb them into the coefficients
$b_{ij},c_{ij}$, although more accurate rate estimates will likely
improve the coefficient estimates). Our data will be measurements of
the RNA concentrations $x$ at many different cellular steady-states
(which correspond to steady-states of the dynamical system). The
problem is to infer the values of the coefficients $b_{ij},c_{ij}$.



\section*{Linear problem at steady-state}

The key to solving this problem efficiently is to restrict our
attention to steady-states, as proposed by Choi (\citeyear{choi12}). This
restriction allows us to transform a nonlinear ODE fitting problem into
a linear regression problem. A steady-state of the system is one in
which RNA and protein levels are constant: $\frac{dx_i}{dt} = \frac
{dy_i}{dt} = 0$. Steady-states of the system correspond to cell states
with roughly constant gene expression levels, like embyronic stem cell,
skin cell or liver cell. In contrast, an embryonic stem cell in the
process of differentiating is not in steady-state. Perturbed
steady-states are particularly interesting. After a perturbation like
gene knockdown, a cell's gene expression levels are in flux for some
time while they adjust to the change. Eventually, if it is still
viable, the cell may settle to a new steady-state [\citet{Huang05}].
These perturbed steady-states are especially helpful for understanding
gene regulation.

In our model, the steady-state conditions $\frac{dx_i}{dt} = \frac
{dy_i}{dt} = 0$ mean that
\[
0 = \tau_i f_i(y) - \lambda_i^{\mathrm{RNA}}
x_i,\qquad 0 = r_ix_i - \lambda_i^{\mathrm{Protein}}
y_i \quad\implies\quad y_i = \frac{r_i x_i}{\lambda
_i^{\mathrm{Protein}}}.
\]
Defining $\tilde{f}_i(z) = f_i(\frac{r_i}{\lambda_i^{\mathrm{Protein}}} z) $ yields
\[
0 = \tau_i \tilde{f}_i(x) - \lambda_i^{\mathrm{RNA}}x_i.
\]
Absorbing the constants into the coefficients $b_{ij},c_{ij}$ (so that
$\tilde{b}_{ij} =\break b_{ij} \Pi_{k \in S_{ij}} \frac{r_k}{\lambda
_k^{\mathrm{Protein}}}$, $\tilde{c}_{ij} = c_{ij} \Pi_{k \in S_{ij}} \frac
{r_k}{\lambda_k^{\mathrm{Protein}}}$), we obtain the final equation
\[
\tau_i \frac{b_{i0} + \sum_j b_{ij} \Pi_{k \in S_{ij}} x_k}{1 + \sum_j
c_{ij} \Pi_{k \in S_{ij}} x_k} - \gamma_i x_i
= 0
\]
or
\[
\tau_i \biggl(b_{i0} + \sum_j
b_{ij} \Pi_{k \in S_{ij}} x_k\biggr) -
\gamma_i x_i \biggl(1 + \sum
_j c_{ij} \Pi_{k \in S_{ij}} x_k
\biggr) = 0
\]
(by multiplying both sides by the denominator). The last equation is
linear in the coefficients $b_{ij},c_{ij}$!
In order to solve for $b_{ij},c_{ij}$, we will need to collect many
different expression measurements $x$ at both naturally occurring and
perturbed steady-states. Each steady-state measurement will lead to a
different linear equation. These equations can be arranged into a
linear system that we can solve for the coefficients.

\section*{Problem formulation}
Our problem is to find $b_{ij},c_{ij}$ such that
\begin{eqnarray*}
0 &=& \tau_i \biggl(b_{i0} + \sum
_j b_{ij} \Pi_{k \in S_{ij}} x_k^{(m)}
\biggr) \\
&&{}- \gamma_i x_i \biggl(1 + \sum
_j c_{ij} \Pi_{k \in S_{ij}} x_k^{(m)}
\biggr)\qquad \forall m = 1,\ldots, M,
\end{eqnarray*}
given RNA expression data $x^{(m)}$ at many different steady-state
points $m=1,\ldots,M$ and known translation and degradation rates $\tau
_i, \lambda_i^{\mathrm{RNA}}, \lambda_i^{\mathrm{Protein}}$. (The experimental means of
collecting the necessary steady-state expression data will be discussed
in the next section.) We solve a separate problem for each gene $i$,
since the coefficients $b_{ij}, c_{ij}$ in the differential equation
$dx_i/dt = \cdots$ for gene $i$ are independent of the coefficients in
the differential equations for other genes. Since we cannot know ahead
of time which potential regulatory terms $\Pi_{k \in S_{ij}} x_k$ are
actually involved, we include all possible terms up to second-order and
look for sparse $b_{ij},c_{ij}$, intepreting $c_{ij} = 0$ to mean that
term $\Pi_{k \in S_{ij}} x_k$ is not a regulator of gene $i$.

Consider gene 2 in the two-gene example. Suppose we have expression
measurements for a naturally occurring steady-state $(x_1^0, x_2^0)$,
and a perturbed steady-state following gene 1-knockout $(0, x_2^1)$. We
obtain two linear equations in the coefficients $b_{20}, c_{21}$:
\begin{eqnarray*}
\tau b_{20} - \lambda_2 x_2^0
\bigl(1 + c_{21} x_1^0 x_2^0
\bigr) &=& 0\qquad \bigl(\mbox{steady-state } \bigl(x_1^0,
x_2^0\bigr)\bigr),
\\
\tau b_{20} - \lambda_2 x_2^1 &=&
0\qquad \bigl(\mbox{steady-state } \bigl(0, x_2^1\bigr)\bigr).
\end{eqnarray*}
If we knew a priori that complex $x_1 x_2$ was the only regulator of
gene 2, these two equations would allow us to solve for the
coefficients ($b_{20} = \frac{\lambda_2 x_2^1}{\tau}$, $c_{21} = \frac
{x_2^0 - x_2^1}{(x_2^0)^2}$). Typically we do not know the regulators
beforehand, however, and we need to use the data to identify them. That
is, we include all possible terms (up to second-order) in the equations:
\begin{eqnarray*}
&&
\tau\bigl(b_{20} + b_{21}x_1^{(m)}
x_2^{(m)} + b_{22} x_1^{(m)}
+ b_{23} x_2^{(m)}\bigr)\\
&&\quad{} - \lambda_2
x_2^{(m)} \bigl(1 + c_{21}x_1^{(m)}
x_2^{(m)} + c_{22} x_1^{(m)}+
c_{22} x_2^{(m)}\bigr) \\
&&\qquad= 0
\end{eqnarray*}
and estimate sparse coefficients $b_{ij}, c_{ij}$ using several
steady-state measurements $(x_1^{(m)}, x_2^{(m)})$. (We should find
that the recovered coefficients $b_{21}, b_{22},\break b_{23}, c_{22},
c_{23}$ are very close to zero, since the corresponding terms do not
appear in the true equation.)

Temporarily suppressing the superscript $m$ denoting the observation,
we can compactly express the general system above by defining $z_i$ as
the vector with entries $z_{i}(j) = \Pi_{k \in S_{ij}} x_k$ [with the\vadjust{\goodbreak}
convention\vspace*{1pt} that $z_i(0) = 1$, $z_i(j) = x_j$ for $j=1,\ldots,n$], which yields
\[
0 = \tau_i b_i^Tz_i -
\gamma_i z_i(i) c_i^Tz_i
\]
for each observation ($m=1,\ldots, M$). If we form a matrix $G_i$ by
concatenating the row vectors $z_i^{(1)},\ldots, z_i^{(M)}$ and let
$D_i$ be a diagonal matrix with entries $z_i^{(m)}(i),   m =
1,\ldots,M$, we can express this as
\[
\left[\matrix{\tau_i G_i & -
\gamma_i D_i G_i}\right] \left[\matrix{b_i
\cr
c_i}\right] = 0
\]
with the constraints $0 \le b_i \le c_i,   c_i(0) = 1$. Stating the
problem in this form elucidates the required number of steady-state
measurements, $M$. If the linear system above were dense and had no
constraints on the coefficients $b_{ij}, c_{ij}$, and the steady-state
expression vectors were (numerically) linearly independent, then we
would require $M = 2 T_{n,k}$, where $T_{n,k}$ is the number of the
terms in the rational-form polynomial of degree $k$ in $n$ genes ($k=2$
if we include up to second-order regulatory interactions). $T_{n,k}$ is
equal to the number of subsets of $\{1,2,\ldots,n\}$ with $k$ or fewer
elements since each term represents an interaction between $j$ distinct
genes ($0 \le j \le k$), hence, $T_{n,k} = \sum_{j=0}^k { n \choose j }
\le n^k $ for $k \le n$ ($T_{n,2} \le n^2$, e.g.). However, the
constraints reduce the dimension of the solution space [$c_i(0) = 1$
reduces it by 1, while $0 \le b_i \le c_i$ reduces it by up to $n$],
and our algorithm also uses $\ell_1$-regression to search for sparse
solutions, which may allow us to reconstruct the coeffcients from far
fewer measurements than $2 T_{n,k}$.

\section*{Experimental approach}

\begin{figure}

\includegraphics{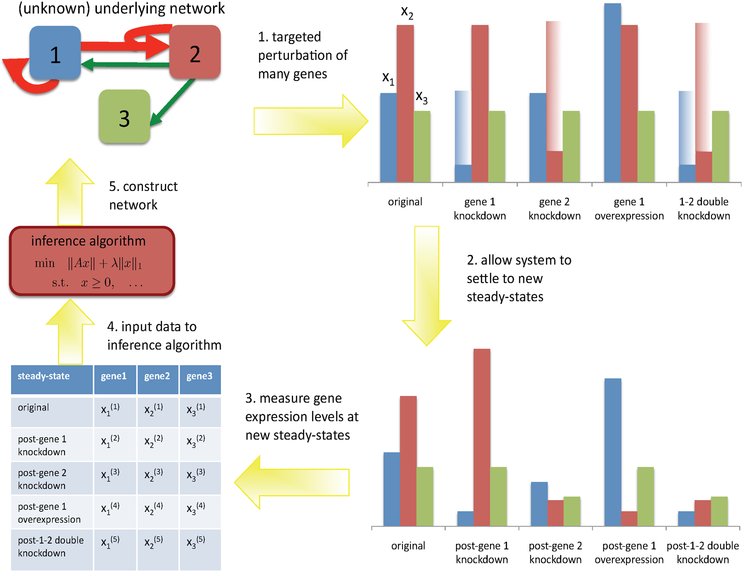}

\caption{Experimental approach for gene network inference. (1) Design
and perform perturbation experiments targeting each gene (or possibly
pair of genes) in the network: these may include overexpression,
knockdowns or knockouts. (2) Following each perturbation, allow the
system to settle to a new steady-state. (3) Measure expression levels
of all genes at each induced steady-state, and collect results in a
data matrix. (4) Use steady-state expression data as input to inference
algorithm. (5) Construct regulatory network from inference algorithm output.}
\label{figexperiments}
\end{figure}

The set of steady-state gene expression measurements needed to fit the
model can be generated via a systematic sequence of gene perturbation
experiments. Figure~\ref{figexperiments} summarizes the overall
approach to finding the regulatory interactions among a set of genes
%
comprising a (roughly) self-contained network of interest. First,
molecular perturbations targeting each gene, or possibly pair of genes,
in the network would be designed and applied one at a time. Following
each perturbation, the cells would be allowed to settle down to a new
steady-state, at which point the gene expression levels would be
measured. The collection of gene expression measurements from different
steady-states would be input to the inference algorithm described in
the next section, which outputs a dynamical systems model of the gene
network capable of predicting the behavior of the network following
other perturbations. Perturbation data not used in the inference
algorithm could be used to validate the recovered model.

The key experimental steps in this procedure, gene perturbations and
gene expression measurements, are established technologies. Gene
perturbations, including overexpression, knockdown and knockout, are
routinely used in biological studies to\vadjust{\goodbreak} investigate gene function.
These experiments can be performed for many laboratory organisms and
cell lines both in vitro and in vivo [\citet{alberts07}].
Overexpression experiments amplify a gene's expression level, usually
by introducing an extra copy of the gene. Knockdown experiments
typically use RNAi technology: the cell is transfected with a short DNA
sequence, driven by a (possibly inducible) promoter element, that
produces siRNA or shRNA that specifically binds the RNA transcripts of
the gene of interest and triggers degradation. Morpholinos can also be
used for gene knockdown. Gene knockout can be achieved by removing all
or part of a gene to permanently disrupt transcription [\citet
{alberts07}]. Overexpression [\citet{rodriguez07}], knockdown [\citet
{rodriguez07,foygel08}] and knockout [\citet{lengner07}] experiments
have all been performed for the Oct4 gene, which helps maintain the
stem cell steady-state. In some cases,\vadjust{\goodbreak} much of the work is already
done: for example, the \emph{Saccharomyces} Genome Deletion Project has
a nearly complete library of deletion mutants [\citet{winzeler99}].

Techniques for gene expression measurement are also well-established.
Gene expression is usually measured at the transcript level: the RNA
transcripts are extracted and reverse-transcribed into cDNA, which can
be quantified with either RT-qPCR, microarray or sequencing
technologies [\citet{alberts07,mortazavi08}]. Housekeeping gene
expression measurements are used as controls to determine the
expression levels of the genes of interest. The gene perturbations and
subsequent expression measurements required to collect data for our
inference algorithm may be time-consuming due to the large number of
perturbations, but all the experimental techniques are quite standard
and resources like deletion libraries can be extremely helpful.

\section*{Algorithm}
We need to solve the linear system
\[
\left[\matrix{\tau_i G_i & -
\gamma_i D_i G_i}\right] \left[\matrix{b_i
\cr
c_i}\right] = 0
\]
for $b_i, c_i$, subject to the constraints $0 \le b_i \le c_i$, $c_i(0)
= 1$. To account for measurement noise and encourage sparsity in
$b_{i}, c_{i}$ (since we know that each gene has only a few
regulators), we will minimize the $\ell_2$-norm error with $\ell_1$
regularization [\citet{tibs96}], which leads to the convex optimization problem
%
\begin{eqnarray}\label{eqopt}
&&\mbox{minimize } \biggl\| \left[\matrix{\tau_i G_i & -
\gamma_i D_i G_i}\right] \left[\matrix{b_i
\cr
c_i}\right] \biggr\|_2 + \lambda\bigl(\| b_i
\|_1 + \| c_i \|_1\bigr)
\nonumber\\[-8pt]\\[-8pt]
&&\quad\mbox{subject to}\quad 0 \le b_i \le c_i,\qquad
c_i(0) = 1,\nonumber
\end{eqnarray}
where $\lambda$ is a parameter controlling sparsity that we can choose
using cross-validation. Since the problem is convex, it can be solved
very efficiently even for large values of $n$ and $m$.

\section*{Nonidentifiability}

Our model's ability to capture self-regulation is very powerful, but it
also leads to a particular form of nonidentifiability. For certain
forms of the equation, given only steady-state measurements, it can be
impossible to determine whether self-regulation is either completely
absent or present in every term. Specifically, any valid equation of
the form
%
\begin{equation}
\label{eqsimple} \frac{dx_i}{dt} = \frac{b_{i0} + \sum_{j=1}^N b_{ij} \Pi_{k \in
S_{ij}} x_k}{1 + \sum_{j=1}^N c_{ij} \Pi_{k \in S_{ij}} x_k} - \gamma_i
x_i,\qquad b_{i0} < 1,
\end{equation}
is indistinguishable at steady-state from any member of the following
family of valid equations indexed by the constant $w$:
%
\begin{equation}
\label{eqhighorder}\quad \frac{dx_i}{dt} = \frac{ (wb_{i0} + \gamma_i)x_i + \sum_{j=1}^N
wb_{ij} \Pi_{k \in{S}_{ij}} x_ix_k}{1 + wx_i + \sum_{j=1}^N wc_{ij} \Pi
_{k \in{S}_{ij}} x_ix_k} -
\gamma_i x_i,\qquad w \ge\frac{\gamma}{1 - b_{i0}}.
\end{equation}
We will refer to these as the ``simple'' and ``higher-order'' forms of the
equation, respectively.
The short proof of their equivalence is given in section S1 %
of the supplementary article [\citet{meister13supplement}]. The
condition $w \ge\frac{\gamma}{1 - b_{i0}}$ guarantees that $w > 0$ and
$0 \le wb_{i0} + \gamma_i \le w$ (since $0 \le b_{i0} < 1$) and $0 \le
wb_{ij} \le wc_{ij}$ (since $0 \le b_{ij} \le c_{ij}$).

We can distinguish between these two alternative forms by measuring the
derivative of the
concentration away from steady-state and comparing it to the derivative
predicted by each form of
the equation. This requires only a few extra thoughtfully-selected
measurements. The details are
in section S2 
of the supplement.

\section*{Simulated six-gene subnetwork in mouse ESC}

\begin{figure}

\includegraphics{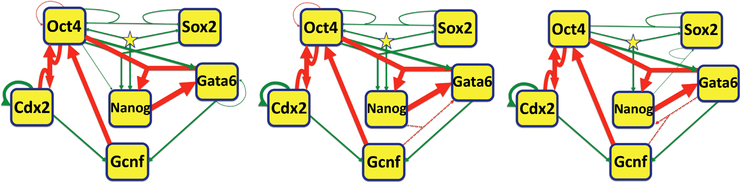}

\caption{Recovery of a synthetic gene regulatory network based on the
biological ESC network using our inference algorithm. The diagrams
represent sytems of ODEs that quantitatively model the gene
interactions. Edge color indicates activation (green) or repression
(red), and edge weights correspond to coefficient magnitudes. The
arrows point from regulator to target, and self-loops indicate
self-regulation. The yellow star represents the third-order complex
OSN. (In addition to all possible first- and second-order terms, we
allow this special third-order term with a free coefficient.) The left
figure represents the original system of ODEs used to generate the
data. The center figure shows the network recovered using our inference
algorithm on noiseless data, and the right figure shows the recovery
with $1\%$ noise added. Both recovered networks reflect coefficient
thresholding at $0.1 \%$ (noiseless case) or $1 \%$ (noisy case) of the
largest recovered coefficent in each gene equation (with the exception
of the noiseless-case Oct4 equation, thresholded at $0.01\%$ to show
the successful recovery of weak edges). The algorithm performs almost
perfectly in the noiseless case, except for a false positive repressor
on Gata6 and two very weak activation edges missing. In the noisy case,
the algorithm recovers all of the strong edges, but misses some of the
weaker ones and returns a few small false positives at our chosen
thresholding level. Overall, the method captures the major network
structure even in the noisy case.}
\label{fignetworks}
\end{figure}

To demonstrate the inference approach, we apply our method to a
synthetic six-gene system based on the Oct4, Sox2, Nanog, Cdx2, Gcnf,
Gata6 subnetwork in a mouse embryonic stem cell (ESC).
\citet{chickarmane08} developed this system based on a synthesis
of knowledge about ESC gene regulation accumulated over the past two
decades [\citet {chickarmane08}]. The network structure is shown
in Figure~\ref{fignetworks}(a), and the detailed model is given by the
following system of ODEs in the six genes:
%
\begin{eqnarray}
\label{eqESC}
\frac{d[O]}{dt} &=& \bigl(0.001 + [A] + 0.005 [O][S] + 0.025
[O][S][N]\bigr)\nonumber\\
&&{}/\bigl(1 + [A] + 0.001[O] + 0.005[O][S]\nonumber\\
&&\hspace*{8.2pt}{} + 0.025[O][S][N] + 10
[O][C] + 10[Gc]\bigr)\nonumber\\
&&{} - 0.1 [O],
\nonumber\\
\frac{d[S]}{dt} &=& \frac{0.001 + 0.005 [O][S] + 0.025 [O][S][N]
}{1 + 0.001[O] + 0.005[O][S] + 0.025[O][S][N]}\nonumber\\
&&{} - 0.1 [S],
\\
\frac{d[N]}{dt} &=& \frac{0.001 + 0.1[O][S] + 0.1 [O][S][N] }{1 +
0.001[O] + 0.1[O][S] + 0.1[O][S][N] + 10[O][G]}\nonumber\\
&&{} - 0.1 [N],
\nonumber\\
\frac{d[C]}{dt} &=& \frac{ 0.001 + 2[C] }{1 + 2[C] + 5[O][C]} - 0.1 [C],
\nonumber\\
\frac{d[Gc]}{dt} &=& \frac{ 0.001 + 0.1[C] + 0.1[G] }{ 1 + 0.1[C]
+ 0.1[G]} - 0.1 [Gc],
\nonumber\\
\frac{d[G]}{dt} &=& \frac{ 0.1 + [O] + 0.00025[G]}{1 + [O] + 0.00025[G]
+ 15[N]} - 0.1 [G].\nonumber
\end{eqnarray}
This model has many of the same qualitative characteristics as the
biological mouse ESC network [\citet{chickarmane08}]. In particular, the
system can support four different steady-states: embryonic stem cell
(ESC), differentiated stem cell (DSC), endoderm and trophectoderm, and
can switch from one to another when certain genes' expression levels
are changed. In the Oct4 equation, $A$~represents an external
activating factor whose concentration $[A]$ depends on the culture
condition. Each of the four steady-states has a corresponding value of
$[A]$: $10$ for ESC and DSC, $25$ for endoderm, and $1$ for
trophectoderm. For the remainder of this paper, we will regard $[A]$ as
known. The explicit system of ODEs (\ref{eqESC}) allows us to generate
data to fit our model and also to quantitatively compare our recovered
solution to the ground truth. The qualitative similarity of this
synthetic network to a real biological network gives us confidence that
our results in this numerical experiment are likely to translate well
to real biological networks.

We observe that the Cdx2, Gcnf and Gata6 equations have alternative
forms (provided we ignore the very small constant term in the $\frac
{d[C]}{dt}$ equation and $[G]$ term in the $\frac{d[G]}{dt}$). With the
minimum possible value of $w$, the alternative
forms are as follows:
%
\begin{eqnarray}\label{eqaltESC}
\frac{d[C]}{dt} &=& \frac{ 0.95 }{1 + 2.5[O] }\qquad (w = 2),
\nonumber\\
\frac{d[Gc]}{dt} &=& \frac{ 0.1001[Gc] + 0.01[C][Gc] +
0.01[Gc][G] }{ 1 + 0.1[Gc] + 0.01[C][Gc] + 0.01[Gc][G]}\nonumber\\
&&{} - 0.1 [Gc] \qquad(w =
0.1),
\\
\frac{d[G]}{dt} &=& \frac{0.111[G] + 0.111[O][G] }{1 + 0.111[G] +
0.111[O][G] + 1.67[N][G]} \nonumber\\
&&{}- 0.1 [G] \qquad(w = 0.111).\nonumber
\end{eqnarray}
To resolve the specific form, we will apply our method twice, once
allowing self-regulation and again disallowing it. Then we will compare
the two recovered forms of each equation and the quality of the fits to
determine whether nonidentifiability exists in each case. If so, we
will break the tie by examining derivatives.

\begin{figure}

\includegraphics{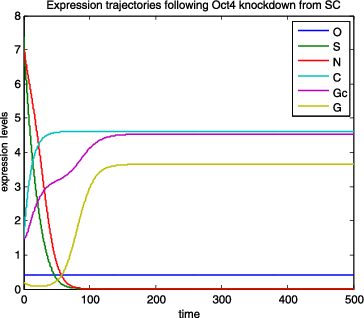}

\caption{Gene expression trajectories during an Oct4 knockdown from SC
steady-state. The expression of Oct4 is artificially reduced to $20\%$
of its SC steady-state expression level and held there, causing the
expression levels of the targets of Oct4 to change in response, which
in turn impact their targets. The system eventually reaches a new
steady-state different from SC. We measure the vector of expression
levels at the new steady-state and use it as data in the inference
algorithm. Since Oct4 is knocked down, this induced steady-state does
not provide useful information about the Oct4 equation, but it is
useful for understanding the role of Oct4 and other genes in the
equations of the remaining five genes.}
\label{figOKD}
\end{figure}

To fit the model, we collect data on the expression levels of all six
genes at many different steady-states. First we measure the expression
levels at all four wildtype steady-states: SC, DSC, endoderm and
trophectoderm. We also induce additional perturbed steady-states by
simulating knockdowns and overexpression of each gene, based on
physical gene perturbation experiments
[\citet{rodriguez07,zafarana09}]. For a knockdown, we hold a gene
at one-fifth of its steady-state expression level; for overexpression
we hold a gene at twice its steady-state level. In each case we wait
for the system to settle to a new steady-state, then measure the
expression levels. Figure~\ref{figOKD} shows the expression
trajectories during Oct4 knockdown from the ESC steady-state as an
example.

%
%

\begin{figure}

\includegraphics{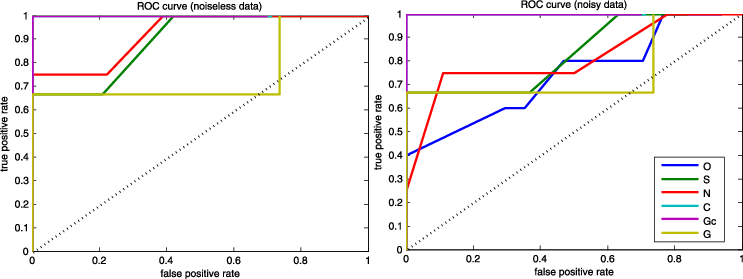}

\caption{ROC curves for recovered networks from noiseless (left) and
noisy (right) data showing the trade-off between true positive rate
(TPR) and false positive rate (FPR) for edge recovery. The ROC curves
show the TPR and FPR that result from a range of coefficient threshold
choices above which we consider an edge to have been recovered. For the
equation $dx_i/dt = \cdots$ and threshold $t$, TPR is defined as the
proportion of true edges $j$ with $c_{ij}^{\mathrm{recovered}} > t$ and FPR
as the proportion of false edges with $c_{ij}^{\mathrm{recovered}} >t$. For
equations with two possible forms, we compare the simple forms of the
true and recovered equations. Each gene equation has a different ROC
curve as indicated by the legend. The dotted black line is the expected
ROC curve for the ``random guessing'' algorithm, while the $(0,1)$ point
corresponds to a perfect algorithm (in fact, our algorithm performs
perfectly for the Gcnf equation).}
\label{figROC}
\end{figure}

The details of the simulation are given in section S3 
of the supplement. We begin by testing the algorithm on noiseless data.
We solve the optimization
problem (\ref{eqopt}) once, then we solve it again with additional
constraints prohibiting
self-regulation. In each case we use cross-validation to select the
sparsity parameter
$\lambda$ (Figure~S1). 
The quality of the fit is comparable for the latter three equations
whether we restrict
self-regulation or not, while for the first three equations restricting
self-regulation has a
significant negative impact on the fit (Table S1), %
indicating that the first three equations are unambiguous while the
last three have two possible
forms. To resolve the nonidentifiability in the latter three equations,
we measure the
derivatives of Cdx3, Gcnf and Gata6 immediately after some additional
informative perturbations:
Oct4, Cdx2 and Nanog knockouts, respectively (Figure S2). %
The test reveals that Gcnf and Gata6 have the simple form, while Cdx2
has a higher-order form. In this example,\ the original coefficients are
recovered almost exactly:
%
\begin{eqnarray}
\label{eqnoiselessrecovery}
\frac{d[O]}{dt} &=& \bigl(0.001 + [A] + \bigl(0.005 [O][S] + 0.025
[O][S][N]\bigr)\bigr)\nonumber\\
&&{}/\bigl(1 + [A] + \bigl(0.001[O] + 0.005[O][S] + 0.025[O][S][N]\bigr)\nonumber\\
&&\hspace*{160.5pt}{} + 10
[O][C] + 10[Gc]\bigr)\nonumber\\
&&{} - 0.1 [O],
\nonumber\\
\frac{d[S]}{dt} &=& \frac{0.001 + 0.005 [O][S] + 0.025 [O][S][N]
}{1 + 0.005[O][S] + 0.025[O][S][N]} - 0.1 [S],
\\
\frac{d[N]}{dt} &=& \frac{0.1 [O][S] + 0.1 [O][S][N] }{1 +
0.1[O][S] + 0.1[O][S][N] + 10[O][G]} - 0.1 [N],
\nonumber\\
\frac{d[C]}{dt} &=& \frac{2[C] }{1 + 2[C] + 5[O][C]} - 0.1 [C],
\nonumber\\
\frac{d[Gc]}{dt} &=& \frac{ 0.001 + 0.1[C] + 0.1[G] }{1 + 0.1[C]
+ 0.1[G]} - 0.1 [Gc],
\nonumber\\
\frac{d[G]}{dt} &=& \frac{ 0.1 + [O]}{1 + [O] + 0.03[N][Gc] + 15[N]} - 0.1
[G].\nonumber
\end{eqnarray}
Next we add zero-mean Gaussian noise to each measurement, with standard
deviation $1 \%$ of the
measurement magnitude. We use the same steady-states as in the
noiseless case, plus
overexpression-knockdown of each pair of genes starting from ESC and
DSC. Using a similar approach
(detailed in section S3 
of the supplement), we recover:
%
\begin{eqnarray} \label{eqnoisyrecovery}
\frac{d[O]}{dt} &=& \frac{ [A] }{1 + [A] + 9.9[Gc] + 9.9 [O][C] } - 0.1
[O],
\nonumber\\[-2pt]
\frac{d[S]}{dt} &=& \frac{0.001 [O][S] + 0.0005[S][N] + 0.025
[O][S][N] }{1 + 0.001[O][S] + 0.0005[S][N] + 0.025[O][S][N]} - 0.1 [S],
\nonumber\\[-2pt]
\frac{d[N]}{dt} &=& \frac{0.09[O][S][N] }{1 + 0.1[G][Gc] +
0.09[O][S][N] + 9.1[O][G]} - 0.1 [N],
\nonumber\\[-9pt]\\[-9pt]
\frac{d[C]}{dt} &=& \frac{ 2[C] }{1 + 2[C] + 5[O][C] } - 0.1 [C],
\nonumber\\[-2pt]
\qquad\frac{d[Gc]}{dt} &=& \frac{ 0.1[C] + 0.1[G] }{ 1 + 0.1[C] +
0.1[G]} - 0.1 [Gc],
\nonumber\\[-2pt]
\frac{d[G]}{dt} &=& \frac{ 0.1 + 0.9[O] }{1 + 0.9[O] + 14.2[N] } - 0.1
[G].\nonumber
\end{eqnarray}
In order to produce clean equations and network diagrams, we choose
appropriate thresholds for each equation below which we zero the
coefficients. (In practice, choosing thresholds is a judgment call
based on the expected number of regulators, the noise level of the data
and the level of detail appropriate for the application.) We set the
thresholds at $0.1\%$ (noiseless case) or $1\%$ (noisy case) of the
largest coefficient recovered for each equation. For example, the
largest recovered coefficient in the $d[G]/dt$ equation is roughly 15
in either case, so we zero the coefficients that fall below $0.015$
(noiseless case) or $0.15$ (noisy case). The recovered systems of
equations shown above reflect these choices. In the noiseless case,
relaxing the threshold on the Oct4 equation to $0.01\%$ leads to the
recovery of more correct terms, listed in parentheses. For
completeness, we also provide receiver operating characteristic (ROC)
curves in Figure~\ref{figROC} to show the trade-off between true
positives and false positives at other thresholds. The network diagrams
in Figure~\ref{fignetworks}(b), (c) include an edge if the corresponding
%
coefficient is above the threshold, with weights reflecting the size of
the coefficients. These diagrams show that the recovery is nearly
perfect in the noiseless case: using the gentler threshold for the Oct4
equation, we recover all the true edges except for three very weak
ones, and return just one small false positive repressor in the Gata6
equation. In the noisy case, we recover all the large coefficients
correctly, although there are a few small false positives and we miss
several of the weakest edges. Overall, the method is able to capture
the major network structure.

\section*{Discussion}

Our experiment on the synthetic ESC system demonstrates that our
algorithm can be used to infer a complex dynamical systems model of
gene regulation and that the method can tolerate low levels of noise.
Term selection from among all possible single gene and gene-complex
regulators (up to second-degree interactions, plus the third-degree
interaction OSN) was successful. The inferred equations are easy to
interpret in terms of gene networks, and the detailed quantitative
information allows for prediction of future expression trajectories
from any starting point.\vadjust{\goodbreak}

The approach is also scalable. Since we have formulated our problem as
a convex optimization problem [equation (\ref{eqopt})], it can be solved
efficiently even for large systems using prepackaged software.
Furthermore, it is trivially parallelizable, since we need to solve a
version of (\ref{eqopt}) to infer the differential equation
coefficients $b_{ij}, c_{ij}$ for each gene $i$. Parallelization is
even more helpful for the cross-validation step, where we need to solve
equation (\ref{eqopt}) for each gene and a sequence of choices sparsity
parameter $\lambda$. We tested the scalability by running the algorithm
with the parallelization discussed above on a simulated 100-gene
system. The algorithm ran correctly in a reasonable time frame (a few
hours) on a computing cluster.

The high resolution of our model is one of its most valuable features,
but it means that accurate term selection may require much data,
especially in the presence of noise. In our experiment, when we added
$1\%$ Gaussian noise, we needed extra data (knockdown/overexpression
pairs) in order to accurately select terms. When we tried $5\%$ noise,
the algorithm consistently selected the large terms in five of the six
equations, but we had to add even more data in order to correctly
identify the major repressor in the Nanog equation. The Nanog equation
is subtle in that Oct4 acts as both an activator in complexes with Sox2
and Nanog and a repressor in a complex with Gata6, so the algorithm
tends to select different Gata6 complexes (or the Gata6 singleton) as
the major repressor when the data is insufficient. In the $5\%$ noise
case, we needed additional data on the role of Gata6 (double-knockdowns
and double-overexpression of pairs including Gata6 from ESC and DSC) in
order to select Oct4-Gata6 as the major repressor of Nanog fairly
consistently. As discussed earlier, another difficulty is the
nonidentifiability that arises from accounting for self-regulation
while restricting data to steady-states. Distinguishing between the two
possible forms of nonidentifiable equations requires extra derivative
data (which can be collected experimentally, although it is more
difficult and time-consuming) and extra steps in the algorithm. The
constraints on the convex optimization problem (\ref{eqopt}), which
arise from thermodynamic considerations, are sufficient to prevent
further nonidentifiability, but in certain cases, certain problems can
suffer from near-nonidentifiability of other forms, which may
contribute to the challenge of term-selection with noisy or limited
data. We ensure accurate term selection by making sure we include
enough diverse, high-quality steady-state measurements.

We should also note that our model does not account for the intrinsic
noise in gene transcription and translation, although these processes
are inherently stochastic, since TF and RNAP binding result from chance
collisions between molecules in the cell. However, the stochastic
version of our rational-form transcription model is highly complex and
there is currently no satisfactory method for its inference. Studying
the deterministic evolution plus additive noise is standard practice
for all but linear models of gene expression, and treatment of the
deterministic model provides insight into the stochastic model. Here we
focus on the additive noise case and leave the study of intrinsic noise
for future investigation.

\section*{Conclusions}

The model we use is based on the detailed thermodynamics of gene
transcription, and quantitatively captures the full spectrum of
regulatory phenomena in a detailed, physically interpretable,
predictive manner. Since we can formulate the model fitting problem as
a convex optimization problem, we can solve it efficiently and scalably
using prepackaged software. $\ell_1$-regularization allows for
term-selection while maintaining the problem convexity. The experiments
required to collect the necessary steady-state gene expression data are
straightforward to perform, as technologies for knockdowns and
overexpression are well-established and measuring gene expression is
relatively simple. The model accounts for activation and repression by
single-protein TFs and synergistic complexes as well as
self-regulation, and describes the magnitude of each type of regulation
in quantitative detail. Furthermore, the model can be extended to
account for environmental effects and auxiliary proteins involved in
regulation, including enhancers and chromatin remodelers. The fitted
model can predict the evolution of the system from any starting point.
Given a set of steady-states gene expression measurements, our
algorithm can be used to fit a model which not only predicts further
steady-states of the system, but also fully describes the transitions
between them. Finally, beside the study of gene regulation, our
approach will be useful in many other application areas where it is
necessary to infer a nonlinear dynamical system by suitable
experimentation and statistical analysis.

\begin{appendix}\label{app}
\section*{Appendix: Thermodynamic model}

In (\ref{eqexpression}), the function $f_i(y)$ represents the
probability that RNAP binds to the $i$th gene promoter. We claim that
$f_i(y)$ has the form
\[
f_i(y) \equiv p_{\mathrm{bound}}^{(i)}(y) =
\frac{ \sum_j e^{-\beta\Delta
\varepsilon_{ij}^{\mathrm{RNAP}}} P e^{-\beta\Delta\varepsilon_{ij}} \Pi_{k \in
S_{ij}} y_k} {
\sum_j (1+e^{-\beta\Delta\varepsilon_{ij}^{\mathrm{RNAP}}} P ) e^{-\beta
\Delta\varepsilon_{ij}} \Pi_{k \in S_{ij}} y_k},
\]
where $\Delta\varepsilon_{ij}$ is the binding energy of the $j$th complex
to the promoter, $\Delta\varepsilon_{ij}^{\mathrm{RNAP}}$ is the binding
energy of RNAP to the $j$th promoter-bound complex, and $P, x_j$ are
the concentrations of RNAP and gene product $j$ [Bintu et~al.
(\citeyear{bintu05apps,bintu05models})].

Any type of regulator (including no regulator at all) can be
represented in this framework. For no regulator, we take $S_{ij} =
\varnothing$ with the convention that $\Pi_{k \in\varnothing} y_k = 1$,
set $\Delta\varepsilon_{ij} =0$, and take $\Delta\varepsilon_{ij}^
{\mathrm{RNAP}}$ as the base binding energy of RNAP to the promoter. For a
repressor, $\Delta\varepsilon_{ij} < 0$ and $\Delta\varepsilon_{ij}^
{\mathrm{RNAP}} > 0$; for an activator, $\Delta\varepsilon_{ij} < 0$ and $\Delta
\varepsilon_{ij}^{\mathrm{RNAP}} < 0$.

Setting
\begin{eqnarray*}
b_{ij} &=& e^{-\beta\Delta\varepsilon_{ij}^{\mathrm{RNAP}}} P e^{-\beta\Delta
\varepsilon_{ij}},
\\
c_{ij} &=& \bigl(1+e^{-\beta\Delta\varepsilon_{ij}^{\mathrm{RNAP}}} P \bigr) e^{-\beta
\Delta\varepsilon_{ij}},
\end{eqnarray*}
we obtain the form given in Section~1:
\[
f_i(y) = \frac{ b_{ij} \Pi_{k \in S_{ij}} y_k } { \sum_j c_{ij} \Pi_{k
\in S_{ij}} y_k }.
\]
Constant terms in the numerator and denominator correspond to the
no-regulator case. Letting $c_{i0}$ denote the constant appearing in
the denominator, our convention will be to divide all of the
coefficients in the numerator and denominator by $c_{i0}$ so that the
constant 1 appears in the denominator.

\subsection*{Simplified derivation}
The derivation we present here follows Bintu et al. and Garcia et al.
[Bintu et~al. (\citeyear{bintu05apps,bintu05models}),
\citet{garcia11}]. For simplicity, we will prove the following
claim for the simplified case with one regulator $y_1$ (as well as the
possibility of RNAP binding with no regulator):
\[
p_{\mathrm{bound}}^{(i)} = \frac{ e^{-\beta\Delta\varepsilon_{i0}^
{\mathrm{RNAP}}} p + e^{-\beta\Delta\varepsilon_{i1}^{\mathrm{RNAP}}} p e^{-\beta
\Delta\varepsilon_{i1}} y_1 } {
(1 + e^{-\beta\Delta\varepsilon_{i0}^{\mathrm{RNAP}}} p) + (1+e^{-\beta
\Delta\varepsilon_{ij}^{\mathrm{RNAP}}} p ) e^{-\beta\Delta\varepsilon_{i1}}
y_1 }.\vadjust{\goodbreak}
\]

We will use the following notation:
$\varepsilon_{P,i1}^S$ is the energy of the state in which RNAP is
specifically bound to the regulator-promoter complex, $\varepsilon
_{P,i0}^S$ is the energy of the state in which RNAP is specifically
bound to the promoter without the regulator,
$\varepsilon_{P}^{\mathrm{NS}}$ is the energy when RNAP is bound to a nonspecific
binding site, $\varepsilon_{i1}^S$ is the energy when $y_1$ is
specifically bound to the promoter, and $\varepsilon_{i1}^{\mathrm{NS}}$ is energy
when $y_1$ is bound to a nonspecific binding site. Then
\begin{eqnarray*}
\Delta\varepsilon_{i0}^{\mathrm{RNAP}} &=& \Delta\varepsilon_{P,i0}
\equiv \varepsilon_{P,i0}^S - \varepsilon_{P}^{\mathrm{NS}},\\
\Delta\varepsilon_{i1}^{\mathrm{RNAP}} &=& \Delta\varepsilon_{P,i1}
\equiv \varepsilon_{P,i1}^S - \varepsilon_{P}^{\mathrm{NS}},\qquad
\Delta\varepsilon_{i1} \equiv\varepsilon_{y_1}^{S} -
\varepsilon_{y_1}^{\mathrm{NS}}.
\end{eqnarray*}
Suppose that we have $j$ RNA polymerase molecules and $k$ molecules of
gene product $1$ (the regulator). We model the genome as a
``reservoir''
with $n$ nonspecific binding sites (to which either RNAP or regulator
can bind). One of these sites is the promoter of gene $i$. Four
different classes of configurations interest us:
\begin{enumerate}
\item empty promoter,
\item regulator bound to promoter,
\item regulator and RNAP bound to promoter,
\item RNAP only bound to promoter.
\end{enumerate}
These correspond to the following partial partition functions, which
represent the ``unnormalized probabilities'' of each configuration:
\begin{enumerate}
\item$Z(j,k)$,
\item$Z(j,k-1)e^{-\beta\varepsilon_{i1}^S}$,
\item$Z(j-1,k-1)e^{-\beta\varepsilon_{i1}^S}
e^{-\beta\varepsilon_{P,i1}^S}$,
\item$Z(j-1,k) e^{-\beta\varepsilon_{P,i0}^S}$,
\end{enumerate}
where $Z(j,k) = \frac{n!}{j!k!(n-j-k)!} e^{-\beta r \varepsilon_{i1}^{\mathrm{NS}}}
e^{-\beta\varepsilon_{P}^{\mathrm{NS}}}$.

$Z(j,k)$ is equal to the total number of arragements of RNAP and
regulator on the nonspecific binding sites times the Boltzmann factor,
which gives the relative probability $e^{-\beta\varepsilon}$ of a
particular state in terms of its energy $\varepsilon$.

Since RNAP binds the promoter only in the third and fourth classes of
configurations, the probability that RNAP binds the promoter is equal
to the unnormalized probability of the third and fourth configurations
divided by the ``total probability'' (the sum of the unnormalized
probabilities of all classes of configurations). Hence,
\begin{eqnarray*}
p_{\mathrm{bound}} & = & \bigl(Z(j-1,k) e^{-\beta\varepsilon_{P,i0}^S} + Z(j-1,k-1)e^{-\beta
\varepsilon_{i1}^S} e^{-\beta\varepsilon_{P,i1}^S}\bigr)\\
&&{}/ \bigl(
Z(j,k) + Z(j-1,k) e^{-\beta\varepsilon_{P,i0}^S}\\
&&\hspace*{6.8pt}{}  + Z(j,k-1)e^{-\beta
\varepsilon_{i1}^S}+ Z(j-1,k-1)e^{-\beta\varepsilon_{i1}^S} e^{-\beta
\varepsilon_{P,i1}^S} \bigr)
\\
&\approx&\biggl(\frac{n^{j-1}n^{k}}{(j-1)!k!} e^{-\beta k \varepsilon
_{i1}^{\mathrm{NS}}} e^{-\beta(j-1) \varepsilon_{P}^{\mathrm{NS}}}e^{-\beta\varepsilon
_{P,i0}^S}\\
&&\hspace*{5pt}{} + \frac{n^{j-1}n^{k-1}}{(j-1)!(k-1)!} e^{-\beta(k-1)
\varepsilon_{i1}^{\mathrm{NS}}} e^{-\beta(j-1) \varepsilon_{P}^{\mathrm{NS}}} e^{-\beta
\varepsilon_{i1}^S} e^{-\beta\varepsilon_{P,i1}^S}\biggr)\\
&&{}\bigg/
\biggl(\frac{n^{j}n^{k}}{j!k!} e^{-\beta k \varepsilon_{i1}^{\mathrm{NS}}} e^{-\beta j
\varepsilon_{P}^{\mathrm{NS}}}\\
&&\hspace*{16pt}{}
+ \frac{n^{j-1}n^{k}}{(j-1)!k!} e^{-\beta k \varepsilon_{i1}^{\mathrm{NS}}}
e^{-\beta(j-1) \varepsilon_{P}^{\mathrm{NS}}}e^{-\beta\varepsilon_{P,i0}^S} + \cdots\biggr)
\\
& =&
\biggl(\frac{j}{n}e^{\beta\varepsilon_{P}^{\mathrm{NS}}}e^{-\beta\varepsilon
_{P,i0}^S} + \frac{j}{n} \frac{k}{n} e^{\beta\varepsilon_{i1}^{\mathrm{NS}}}
e^{\beta\varepsilon_{P}^{\mathrm{NS}}} e^{-\beta\varepsilon_{i1}^S} e^{-\beta
\varepsilon_{P,i1}^S}\biggr)\\
&&{}\bigg/\biggl(1 + \frac{j}{n}e^{\beta\varepsilon_{P}^{\mathrm{NS}}}e^{-\beta\varepsilon
_{P,i0}^S} + \frac{k}{n} e^{\beta\varepsilon_{i1}^{\mathrm{NS}}}
e^{-\beta\varepsilon_{i1}^S}\\
&&\hspace*{37pt}{}+ \frac{j}{n} \frac{k}{n} e^{\beta\varepsilon_{i1}^{\mathrm{NS}}} e^{\beta\varepsilon
_{P}^{\mathrm{NS}}} e^{-\beta\varepsilon_{i1}^S} e^{-\beta\varepsilon_{P,i1}^S}\biggr)
\\
& = & \frac{ ({j}/{n}) e^{-\beta\Delta\varepsilon_{P,i0}} + ({j}/{n})
({k}/{n}) e^{ -\beta\Delta\varepsilon_{i1}} e^{-\beta\Delta\varepsilon
_{P,i1}} } {
1 + ({j}/{n}) e^{-\beta\Delta\varepsilon_{P,i0}}
+ ({k}/{n}) e^{ -\beta\Delta\varepsilon_{i1}}
+ ({j}/{n}) ({k}/{n}) e^{ -\beta\Delta\varepsilon_{i1}} e^{-\beta
\Delta\varepsilon_{P,i1}} }
\\
& = & \frac{ ({j}/{n}) e^{-\beta\Delta\varepsilon_{P,i0}} + ({j}/{n})
({k}/{n}) e^{ -\beta\Delta\varepsilon_{i1}} e^{-\beta\Delta\varepsilon
_{P,i1}} } {
1 + ({j}/{n}) e^{-\beta\Delta\varepsilon_{P,i0}} + ({k}/{n}) e^{
-\beta\Delta\varepsilon_{i1}} ( 1 + ({j}/{n}) e^{-\beta\Delta\varepsilon
_{P,i1}} ) }
\\
& = & \frac{ p e^{-\beta\Delta\varepsilon_{i0}^{\mathrm{RNAP}}} + p y_1 e^{
-\beta\Delta\varepsilon_{i1}} e^{-\beta\Delta\varepsilon_{i1}^
{\mathrm{RNAP}}} } {
1 + p e^{-\beta\Delta\varepsilon_{i0}^{\mathrm{RNAP}}} + y_1 e^{ -\beta
\Delta\varepsilon_{i1}} ( 1 + p e^{-\beta\Delta\varepsilon_{i1}^
{\mathrm{RNAP}}} ) },
\end{eqnarray*}
where in the second line we used the approximation $\frac
{n!}{j!k!(n-j-k)!} \approx\frac{n^{j}n^{k}}{j!k!} $ which holds for
$j, k \ll n$, in the third we divided by $\frac{n^{j}n^{k}}{j!k!}
e^{-\beta k \varepsilon_{i1}^{\mathrm{NS}}}e^{-\beta\varepsilon_{P}^{\mathrm{NS}}}$, in the
fourth we used the identities $\Delta\varepsilon_{P,i0} = \varepsilon
_{P,i0}^S - \varepsilon_{P}^{\mathrm{NS}}$, $\Delta\varepsilon_{P,i1} = \varepsilon
_{P,i1}^S - \varepsilon_{P}^{\mathrm{NS}}$, $\Delta\varepsilon_{i1} \equiv\varepsilon
_{i1}^{S} - \varepsilon_{i1}^{\mathrm{NS}}$, and in the last we substituted in the
definitions $\frac{j}{n} = p$, $\frac{k}{n} = y_1$, $\Delta\varepsilon
_{i0}^{\mathrm{RNAP}} = \Delta\varepsilon_{P,i0}$, $\Delta\varepsilon_{i1}^
{\mathrm{RNAP}} = \Delta\varepsilon_{P,i1}$.
\end{appendix}

\section*{Acknowledgments}

Many thanks to Xi Chen for helpful discussions.

\begin{supplement}
\stitle{Nonidentifiability, tie-breaking and synthetic network study details}
\slink[doi]{10.1214/13-AOAS645SUPP} 
\sdatatype{.pdf}
\sfilename{aoas465\_supp.pdf}
\sdescription{We discuss nonidentifiability and tie-breaking in
Sections S1
and S2 
by proving the equivalence of two different equation forms at
steady-state and describing methods
for determining the true form of an ambiguous equation. In Section S3 %
we provide the details of our study of a simulated six-gene network in
mouse ESC, including parameter selection, tie-breaking and
thresholding.}
\end{supplement}


\printaddresses

\end{document}